\documentclass[%
 reprint,
 amsmath,amssymb,
 aps,
prc,
]{revtex4-2}

\usepackage{graphicx}
\usepackage{dcolumn}
\usepackage{bm}
\usepackage[utf8]{inputenc}
\usepackage{textgreek}
\usepackage{amsmath}
\usepackage{natbib}
\usepackage{xcolor}
\usepackage{comment}
\begin{document}

\preprint{APS/123-QED}

\title{Non-Equilibrium Trace Anomaly And Bulk Viscosity in Heavy Ion Collisions From Kinetic Theory}

\author{Krishanu Sengupta}
\email{krishanu.sengupta@niser.ac.in}
\affiliation{School of Physical Sciences, National Institute of Science Education and Research, An OCC of Homi Bhabha National Institute, Jatni - 752050, India}
\author{Reghukrishnan Gangadharan}%
 \email{reghukrishnang@niser.ac.in}
\affiliation{School of Physical Sciences, National Institute of Science Education and Research, An OCC of Homi Bhabha National Institute, Jatni - 752050, India}
\author{Victor Roy}
 \email{victor@niser.ac.in}
\affiliation{School of Physical Sciences, National Institute of Science Education and Research, An OCC of Homi Bhabha National Institute, Jatni - 752050, India}
\date{\today}

\begin{abstract}
We investigate the far-from-equilibrium dynamics and transport properties of a relativistic massive gas obeying Maxwell-Boltzmann (MB), Bose–Einstein (BE), and Fermi–Dirac (FD) statistics, undergoing a boost-invariant Bjorken expansion. We solve the relativistic Boltzmann equation in the relaxation-time approximation (RTA) using the method of moments. 
We focus specifically on the time evolution of the trace of the energy-momentum tensor $\Theta_{\; \mu}^{\mu}$ and the bulk viscous pressure $\Pi$, quantities that are crucial for understanding the breaking of conformal symmetry in the rapidly evolving fireball created in heavy ion collisions. Our results demonstrate that the non-equilibrium values of $\Theta^{\mu}_{\; \mu}/T^{4}$ exhibits non-monotonic behavior, characterized by a local maximum at early times and a distinct dip around the characteristic collision time scale $\tau_R$. We found the scaled bulk pressure $\Pi/P_{0}$ ($P_{0}$ isotropic pressure) depends significantly on the particle statistics. Along with this, we also observe that the magnitudes of both the bulk and $\Theta^{\mu}_{\; \mu}/T^{4}$ are significantly enhanced by a larger initial chemical potential. Furthermore, by initializing the system with random non-equilibrium configurations, we show that the evolution trajectories of the scaled bulk pressure and pressure anisotropy converge to a common late-time solution. 


\end{abstract}

\maketitle


\section{\label{sec:Intro}Introduction}
Heavy ion collisions at relativistic energies create extreme conditions of nuclear-matter where quarks and gluons become asymptotically free over the nuclear volume for a short period of time and form a novel state of matter called  quark-gluon   (QGP) \cite{PhysRevLett.30.1343, DAVIDPOLITZER1974129, PhysRevLett.34.1353,RevModPhys.89.035001,Pasechnik_2017}. These collision experiments at RHIC and LHC facilities provide us with unique insights into the fundamental properties of quantum chromodynamics (QCD) and the nature of the QGP \cite{ARSENE20051, BACK200528, ADCOX2005184, ADAMS2005102, SHURYAK200564, Muller:2012zq, FOKA2016172}. 

Among the various thermodynamic quantities that characterize the QGP, the trace anomaly, defined as the deviation of the trace of the energy-momentum tensor $g^{\mu\nu}T_{\mu\nu}$ from its conformal value, plays an important role in understanding the equation of state and transport properties of strongly interacting matter \cite{PhysRevD.16.438, PhysRevD.80.014504,PhysRevD.97.014510, Kharzeev:2007wb, PhysRevD.84.087703}. The trace anomaly is given by $\Theta_{\;\mu}^{\mu} = \varepsilon - 3P$, where $\varepsilon$ is the energy density and $P$ is the pressure. In classical field theories conformal symmetry (scale invariance) implies the trace of the energy-momentum tensor should vanish \(\Theta_{\; \mu}^{\mu}=g^{\mu\nu}T_{\mu\nu}=0\). The trace anomaly is sometimes also referred to as the interaction measure and in Quantum Chromo Dynamics (QCD) it becomes non zero due to the quantum corrections \cite{BAZAVOV2014867, PhysRevD.90.094503}. 

However, calling the trace anomaly an interaction measure is somewhat of a misnomer since in strongly interacting conformal matter where $\varepsilon = 3P$, the trace anomaly vanishes, yet the interactions remain strong. Rather, a large value of the trace anomaly indicates a significant deviation from conformality. The trace anomaly is also closely connected to the bulk viscosity, which governs the system's response to compression and expansion. The bulk viscosity can be extracted from the two-point correlation function of the trace anomaly through a Kubo formula~ \cite{PhysRevLett.100.162001, Kharzeev:2007wb, KARSCH2008217,PhysRevD.80.034023, PhysRevD.84.087703}. 
\begin{equation}
\zeta = \lim_{\omega \to 0} \frac{1}{9\omega T} \int_0^\infty dt \, e^{i\omega t} \langle \Theta(t) \Theta(0) \rangle,
\end{equation}
where $T$ is the temperature.
Meyer showed that lattice calculations of this trace-anomaly correlator at finite temperature provide direct access to the bulk viscosity in the continuum limit~\cite{PhysRevLett.100.162001}.

While lattice QCD calculations provide rigorous results for the trace anomaly and bulk viscosity in equilibrium matter, the study of non-equilibrium dynamics requires different approaches. Lattice studies predict that the equilibrium trace anomaly exhibits a pronounced peak near the crossover temperature $T_c \sim 155$ MeV, where the transition from QGP to hadron gas occurs, signaling strong breaking of conformal symmetry ~\cite{PhysRevLett.100.162001, Ding:2015ona,DING201452, PhysRevLett.113.082001}. However, in the rapidly evolving fireball created in heavy ion collisions, the system is far from equilibrium, and its dynamics must be described using kinetic theory. The method of moments, pioneered by Harald Grad in 1950's and later used by Isreal and Stewarts in their seminal paper \cite{1979AnPhy.118..341I}. Modern developments began by Denicol et al.~\cite{PhysRevD.85.114047, PhysRevLett.105.162501,PhysRevLett.106.212302} and further developed by Speranza and Romatschke, provides a powerful framework to study such non-equilibrium systems through the relativistic Boltzmann equation. In this approach, the energy density and pressure deviate from their equilibrium relation due to large gradients and finite relaxation times which results into non-equilibrium evolution. Consequently, the trace of the energy-momentum tensor $\Theta_{\mu}^{\mu} = \varepsilon - 3P$ receives contributions from both the equilibrium breaking of conformal symmetry and the non-equilibrium bulk viscous dynamics. This interplay between equilibrium and dynamical effects is important for understanding the transport properties of the evolving QGP.

In this study we investigate the trace anomaly and hence the bulk viscosity for the rapidly expanding fireball in heavy ion collisions using kinetic theory and moments of single particle distribution function for a massive gas.


Recently, the study of non-equilibrium dynamics and hydrodynamic attractors in heavy-ion collisions has gained a lot of traction after the seminal work published by Spalinsky and Heller \cite{PhysRevLett.125.132301, PhysRevD.97.036020, SPALINSKI2018468, PhysRevLett.120.012301, Strickland:2018ayk, BLAIZOT2021136478, JANKOWSKI2023104048,   PhysRevD.109.074020, DASH2020135481, Soloviev:2021lhs}. Recent developments also exhibited the importance of non-equilibrium effects in determining transport coefficients and understanding the approach to hydrodynamic behavior through the concept of hydrodynamic attractors for various types of system including massless conformal and massive cases \cite{2017JHEP...12..079R, PhysRevLett.124.152301, CHATTOPADHYAY2022136820,PhysRevC.105.024911}. The direct computation of time-dependent moments of the single-particle distribution function, including bulk pressure and viscous coefficients, without the approximations inherent in truncated hydrodynamic theories
allow us to study these quantities far-away from the local equilibrium \cite{FLORKOWSKI2013249, HEINZ2016193, PhysRevD.91.085024, PhysRevC.109.L051901, PhysRevC.90.044908}.

The computation of trace anomaly and bulk viscosity from non-equilibrium attractor solutions represents a shift from traditional approaches. Rather than assuming equilibrium values, this framework allows for the self-consistent determination of these quantities from the evolving distribution function as it approaches the attractor solution which also converges with hydrodynamic results. This approach is particularly relevant for heavy ion collisions, where the system's evolution is dominated by rapid expansion and the approach to local equilibrium occurs on time scales comparable to the collision dynamics.

Throughout our study, we use natural units and set \(\hbar=c=k_B=1\). Our space-time is flat with mostly negative sign metric tensor \(g_{\mu\nu}=\text{diag}(+1,-1,-1,-1)\) .

\subsection{\label{sec:level2} Formulation: massive gas undergoing Bjorken expansion}

To study the far-from-equilibrium evolution of a non-conformal system, we consider a system consisting of massive particles with rest mass $m$ undergoing longitudinal boost-invariant Bjorken flow \cite{PhysRevD.27.140}. The microscopic dynamics of such an out-of-equilibrium system is described by the relativistic Boltzmann transport equation
\begin{equation}
    p^{\mu}\partial_{\mu} f_{\mathbf{p}} - \Gamma^{\rho}_{\mu\nu}\, p^{\mu}p^{\nu}\,\frac{\partial f_{\mathbf{p}}}{\partial p^{\rho}} = C[f],
    \label{eq:boltzmann_general}
\end{equation}
together with the on-shell condition $
    p^{\mu}p_{\mu}=m^{2}$ , the above becomes 
    \begin{equation}
        p^{\mu}\partial_{\mu} f_{\mathbf{p}} - \Gamma^{i}_{\mu\nu}\, p^{\mu}p^{\nu}\,\frac{\partial f_{\mathbf{p}}}{\partial p^{i}} = C[f],
    \end{equation}
where $f_{\mathbf{p}}$ is the single-particle distribution function, $C[f]$ is the collision term, $p^{\mu}=(p^{0},\mathbf{p})$ is the particle four-momentum, and $\Gamma^{\rho}_{\mu\nu}$ are the Christoffel symbols \cite{Cercignani1975, ANDERSON1974466, DenicolNoronha2016}.

In kinetic theory, the conserved particle four-current $N^{\mu}$ and the energy-momentum tensor $T^{\mu\nu}$ are defined as the first and second moments of $f_{\mathbf{p}}$,
\begin{equation}
    N^{\mu} = \int \frac{d^{3}\mathbf{p}}{(2\pi)^{3}p^{0}}\, p^{\mu} f_{\mathbf{p}}, \qquad
    T^{\mu\nu} = \int \frac{d^{3}\mathbf{p}}{(2\pi)^{3}p^{0}}\, p^{\mu}p^{\nu} f_{\mathbf{p}}.
    \label{eq:Nmumu_Tmunu}
\end{equation}
For out-of-equilibrium systems, thermodynamic quantities must be defined through a matching prescription. We employ the Landau matching conditions, which equate the out-of-equilibrium energy density with the equilibrium energy density at the effective temperature $T$ 
\begin{equation} \varepsilon = \int \frac{d^{3}\mathbf{p}}{(2\pi)^3}p^{0}f_{\mathbf{p}} = \int \frac{d^{3}\mathbf{p}}{(2\pi)^3}p^{0}f_{eq}. 
\label{eq:Landau-matching energy}
\end{equation} 
Similarly, matching the number densities determines the chemical potential 
\begin{equation} n = \int \frac{d^3 \mathbf{p}}{(2 \pi)^3}f_{\mathbf{p}} = \int \frac{d^3 \mathbf{p}}{(2 \pi)^3}f_{eq},
\label{eq:Landau-matching number}
\end{equation}
where $f_{eq}$ is the equilibrium distribution function.
\\
To model the collision term, we employ the relaxation time approximation (RTA), for which the collision term $C[f]$ simplifies to
\begin{equation}
    C[f] = -u_{\mu}p^{\mu}\frac{f_{\mathbf{p}}-f_{eq}}{\tau_{R}},
    \label{eq:rta_collision}
\end{equation}
where $\tau_{R}$ is the relaxation time, taken here to be momentum-independent for simplicity \cite{qgy4-bhnq}.
\\
The most convenient coordinate system to describe Bjorken flow is Milne coordinates $x^{\mu}=(\tau,x,y,\eta)$, where $\tau=\sqrt{t^{2}-z^{2}}$ is the longitudinal proper time and $\eta=\tanh^{-1}(z/t)$ is the spacetime rapidity. In Milne coordinates, Bjorken flow corresponds to a fluid at rest,
\begin{equation}
    u^{\mu}=(1,0,0,0),
\end{equation}
and the expansion rate is $\theta=u^{\mu}{}_{;\mu}\sim 1/\tau$. Due to the rapid expansion at early times, the system can be driven far from local thermal equilibrium.

 In this co-ordinate system along with the RTA approximation Eq.~\eqref{eq:boltzmann_general} reduces (for Bjorken symmetry) to
\begin{equation}
    \frac{\partial f_{\mathbf{p}}}{\partial \tau}
    - \frac{p_{z}}{\tau}\frac{\partial f_{\mathbf{p}}}{\partial p_{z}}
    = -\frac{f_{\mathbf{p}}-f_{eq}}{\tau_{R}}.
    \label{eq:boltzmann_bjorken_rta_pz}
\end{equation}


This equilibrium distribution function depends on the effective temperature $T(\tau)$ and chemical potential $\mu(\tau)$ determined from the Landau matching conditions \eqref{eq:Landau-matching energy}, \eqref{eq:Landau-matching number} respectively. In the present boost-invariant setup, $f_{eq}$ depends on the transverse momentum $p_{T}=\sqrt{p_{x}^{2}+p_{y}^{2}}$, the longitudinal momentum $p_{\eta}=t p_{z}-z E_{p}$, and the proper time $\tau$.

Other quantities of interest, such as the shear and bulk stress tensors, can also be obtained from specific projections of second- or higher-order moments \cite{ 10.1093/acprof:oso/9780198528906.001.0001, Denicol:2021Microscopic, Jaiswal:2016hydroReview}. For example, the traceless symmetric shear stress tensor $\pi^{\mu \nu}$ is defined as the spatial projection of the second moment  
\begin{equation}
    \pi^{\mu \nu} = \int\frac{d^3 \mathbf{p}}{ (2 \pi)^3 p^{0}}\Delta^{\mu \nu}_{\; \alpha \beta}  p^{\alpha}p^{\beta}f_{\mathbf{p}},
\end{equation}
where 
$$
\Delta^{\mu \nu}_{\; \alpha \beta} = \frac{1}{2} \bigg( \Delta^{\mu}_{\;\alpha} \Delta^{\nu}_{\; \beta} + \Delta^{\mu}_{\; \beta} \Delta^{\nu}_{\; \alpha} - \frac{2}{3} \Delta^{\mu \nu}\Delta_{\alpha \beta}\bigg),
$$
and $\Delta^{\mu \nu} = g^{\mu \nu} - u^{\mu}u^{\nu}$. The bulk pressure $\Pi$ is defined as 
\begin{equation}
    \Pi +P_{0} = - \frac{1}{3} \int\frac{d^3 \mathbf{p}}{ (2 \pi)^3 p^{0}}\Delta_{\mu \nu} p^{\mu}p^{\nu}f_{\mathbf{p}},
\end{equation}
where $P_{0}$ is the isotropic pressure. For a massless gas in equilibrium, $\Pi = 0$. The total energy-momentum tensor $T^{\mu \nu}$ in Landau frame, decomposed  as ideal plus dissipative correction
\begin{equation}
    T^{\mu \nu} = \varepsilon u^{\mu} u^{\nu} - \Delta^{\mu \nu}P_{0} - \Delta^{\mu \nu}\Pi + \pi^{\mu \nu}.
\end{equation}
first two terms corresponds to ideal and the last two as dissipative correction. 
\\

Based on different types of particles, we know three different types of equilibrium distribution functions they are 
\begin{equation}
    f_{eq}=\left[\exp\!\left(\frac{-p^{0}+\mu}{T}\right)+r\right]^{-1},
    \label{eq:feq_r}
\end{equation}
$r = 0$, corresponds to Maxwell-Boltzmann, $r = 1$ corresponds to Fermi-Dirac and $r = -1$ is for Bose-Einstein.

For massless Boltzmann gas 
\begin{equation} 
\varepsilon = e^{\mu/T}3T^4/\pi^2, 
\end{equation} 
and for massless FD and BE distribution 
\begin{eqnarray}
    \varepsilon = \frac{T^4}{4 \pi^2} \int^{\infty}_{0} \frac{y^3}{\exp{[-\alpha + y] + r}}dy \qquad 
\end{eqnarray}
where $y = p^{0}/T$ and $\alpha = \mu /T$. Finally,
the thermodynamic pressure is given as $P_{0} = \varepsilon/3$.
\\

\textbf{Method of moments:}
We use the relativistic method of moments to study the evolution of the non-equilibrium system. In this approach, we take moments of various orders of the Boltzmann transport equation, assuming the distribution function $f_{\mathbf{p}}$ can be reconstructed from its moments. This procedure transforms the Boltzmann equation into an infinite hierarchy of coupled differential equations for these moments.

The construction of appropriate moments for relativistic kinetic theory has been explored through different formulations in the literature. Notably, Denicol \cite{PhysRevD.85.114047,DenicolNoronha2016} and later Yan and Blaizot ~\cite{BLAIZOT2021136478, PhysRevC.104.055201,BLAIZOT2018283} proposed a moment construction using powers of momentum components that is particularly well-suited for systems with Bjorken symmetry and optimized for studying early-time dynamics in heavy-ion collisions.

Under Bjorken flow, the essential microscopic information of $f_{\mathbf{p}}$ is fully encoded in a specific set of moments. Following the construction suitable for boost-invariant systems~\cite{DenicolNoronha2016}, these moments are defined as 
\begin{equation}
    \rho_{n,l} = \int \frac{d^3 \mathbf{p}}{(2 \pi)^3} (p^{0})^n \left(\frac{p_{z}}{p^{0}} \right)^{2l} f_{\mathbf{p}},
    \label{eq:neqMoment}
\end{equation}
where $n$ characterizes the radial structure in momentum space and $l$ captures the angular anisotropy. The corresponding equilibrium moments $\rho^{eq}_{n,l}$ are obtained by replacing $f_{\mathbf{p}}$ with $f_{eq}$ in the above integral
\begin{equation}
\rho^{eq}_{n,l} = \int \frac{d^3 \mathbf{p}}{(2 \pi)^3} (p^{0})^n \left(\frac{p_{z}}{p^{0}} \right)^{2l} f_{eq}.
\label{eq:eqMoment}
\end{equation}
For massive particles, this can be expressed analytically as 
\begin{equation}
    \rho^{eq}_{n,l} =  \frac{T^{n+3}}{2\pi^{2} (2l+1)} G_{n+2,l,1} (\chi , \alpha , r),
\end{equation}
where the function $G_{n,l,\lambda}(\chi, \alpha, r)$ is defined as 
\begin{equation}
    G_{n,l, \lambda}(\chi, \alpha, r) = \int^{\infty}_{\chi} \frac{(y^{2} - \chi^{2})^{l + 1/2} y^{n-2l-1}}{\bigg[ \exp[- \alpha +y] + r\bigg]^{\lambda}}  dy,
\end{equation}
with $\chi = m/T$. In the massless limit of MB distribution, $G_{n,l,0}(0,0,0) = e^{\alpha} n!$, which gives $\rho^{eq}_{1,0} = 3e^{\mu/T}T^4/\pi^2$.

Substituting Eqs.~\eqref{eq:neqMoment} and \eqref{eq:eqMoment} into Eq.~\eqref{eq:boltzmann_bjorken_rta_pz}, we obtain the evolution equation for the $(n,l)$ moment
\begin{equation}
    \partial_{\tau} \rho_{n,l} + \frac{(2l + 1)}{\tau} \rho_{n,l} + \frac{n - 2l}{\tau} \rho_{n,l+1} = -\frac{1}{\tau_{R}} (\rho_{n,l} - \rho^{eq}_{n,l}),
    \label{eq:r14}
\end{equation}
here $n \in \mathbb{Z}$ and $l = 0,1,2,...$. Clearly, this constitutes an infinite hierarchy of coupled differential equations (coupled via the index $l$) for the non-equilibrium moments, which must be solved by appropriate truncation.

Various moments correspond to different thermodynamic quantities. For example, the energy density and number density are given by the $(n=1,l=0)$ and $(n=0,l=0)$ moments of the distribution
\begin{eqnarray}
\varepsilon &=& \rho_{1,0},\\ 
{\rm n} &=& \rho_{0,0}.
\end{eqnarray}
The longitudinal pressure (along the beam direction) and the isotropic pressure $P_{0}$ are given by the $(n=1,l=1)$ moment of the non-equilibrium distribution and equilibrium distribution repectively
\begin{equation}
    P_{L} = \rho_{1,1}, \qquad P_{0} = \rho^{eq}_{1,1}.
\end{equation}
The transverse pressure $P_T = \frac{3}{2} \left[ P_{0} + \Pi - \frac{1}{3} P_{L} \right]$ can be expressed in terms of moments as
\begin{equation}
    P_{T} = \frac{1}{2}\left(\rho_{1,0}^{eq} - m^2\rho_{-1,0} - \rho_{1,1} \right).
\end{equation}

To derive this expression, we first note that the bulk viscous pressure $\Pi$ is obtained from the trace of the energy-momentum tensor. Using the kinetic theory definition, the trace is
\begin{equation}
    \Theta_{\;\mu}^{\mu} = \int \frac{d^3\mathbf{p}}{(2\pi)^3 p^0} m^2 f_{\mathbf{p}} = m^2 \rho_{-1,0},
\end{equation}
where we have used the on-shell condition $p_\mu p^\mu = m^2$. From the decomposition of $T^{\mu\nu}$ and noting that the shear stress is traceless ($\pi^{\mu}_{\;\mu} = 0$), we have
\begin{equation}
    \Theta_{\;\mu}^{\mu} = \varepsilon - 3(P_{0} + \Pi) = m^2 \rho_{-1,0}.
\end{equation}
Solving for the bulk pressure we have
\begin{equation}
    \Pi = \frac{\varepsilon - 3P_{0} - m^2\rho_{-1,0}}{3} = \frac{1}{3}\bigg(\rho_{1,0} - 3\rho_{1,1}^{eq} - m^2 \rho_{-1,0} \bigg).
\end{equation}

The pressure anisotropy (shear pressure) is characterized by
\begin{equation}
    \phi = \frac{2}{3} \left( P_{T} - P_{L} \right) = \frac{1}{3}\left(\rho^{eq}_{1,0} - m^2\rho_{-1,0} - 3\rho_{1,1}\right).
\end{equation}
In the isotropic limit, $P_T = P_L = P_0$ and both $\Pi$ and $\phi$ vanish, as expected for an equilibrium system.

In kinetic theory the positivity of the distribution function, $f_{\mathbf{p}}\geq 0$, ensures that the effective transverse and longitudinal pressures are non-negative,
$$
P_T=\tfrac12\langle p_T^2\rangle \geq 0,\qquad  
P_L=\frac{\langle p_{\eta}^2\rangle}{\tau^2} \geq 0 .
$$
Thus the total isotropic pressure,
$$
P_{0}+\Pi=\tfrac13\left\langle p_T^2 + \frac{p_{\eta}^2}{\tau^2}\right\rangle ,
$$
is also non-negative.
The trace of the energy-momentum tensor satisfies
$$
T^\mu_{\ \mu}=m^2\langle 1\rangle = \varepsilon - 3(P_{0}+\Pi) \geq 0 ,
$$
providing an additional constraint.
Introducing normalized ,stresses $\pi/P_{0}$ and $\Pi/P_{0}$ (where $\pi^{\eta}_{\eta}\equiv\pi$), these conditions imply
\begin{equation}
\frac{\Pi}{P_{0}} + \frac{\pi}{2P_{0}}  \geq -1,\quad
\frac{\Pi}{P_{0}} - \frac{\pi}{P_{0}} \geq -1,\
\label{eq:Constraint_1}
\end{equation}
\begin{equation}
\frac{\Pi}{P_{0}} \geq -1,\quad
\frac{\Pi}{P_{0}} \leq \frac{\varepsilon}{3P_{0}}-1.
\label{eq:Constraint_2}
\end{equation}
The first three bounds are universal, while the last one depends on the ratio $m/T$, and therefore varies with the temperature of the system
\cite{CHATTOPADHYAY2022136820,PhysRevC.105.024911}.

\subsection{Results}
We now explore the non-equilibrium evolution of the system through the time evolution of the various moments defined above. We consider two different scenarios for system initialization. In the first case, we start from an equilibrium configuration. Due to the rapid expansion during the early time, the system begins deviating from local equilibrium (as defined by the Landau matching conditions) until collisions start dominating around $\tau \sim \tau_{R}$, bringing the system back closer to local equilibrium at late times $\tau \gg \tau_{R}$. The interplay between expansion (which drives the system away from equilibrium) and collisions (which restore equilibrium) is reflected in the temporal evolution of nearly all moments.

For both initial conditions, we start with an initial temperature $T_0 = 1$ GeV and rest mass $m = 0.1$ GeV, giving an initial mass parameter $\chi_0 = m/T_0 = 0.1$. We also set the initial scaled chemical potential to $\alpha_0 = 0$ unless stated otherwise.

\begin{figure}[h]
    \centering
    \hspace*{-2.5mm}\includegraphics[width=1.08\linewidth]{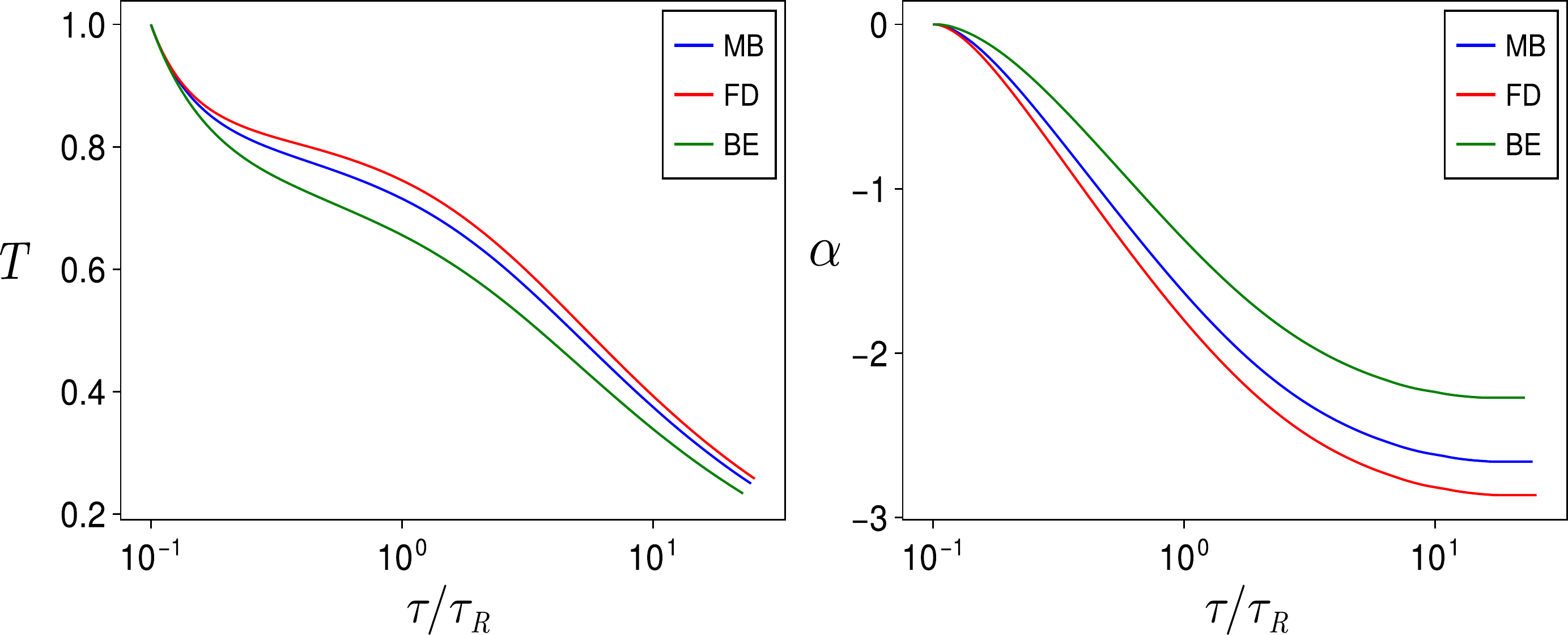}
    \caption{Time evolution of $T$ and $\alpha$ for three different distributions with initial conditions $T_0=1~\mathrm{GeV}$ and $\alpha_0=0$. The $x$-axis is shown on a logarithmic scale.}
    \label{fig:TandMu}
\end{figure}

The evolution of the local temperature and chemical potential are two important inputs to the moment equations. In Fig.~\ref{fig:TandMu}, we show the temporal evolution of $T$ and $\alpha$ as obtained from the Landau matching conditions \eqref{eq:Landau-matching energy}, \eqref{eq:Landau-matching number}, as functions of the scaled time $\tau/\tau_{R}$.

As expected, both the temperature and chemical potential decrease as functions of time, with a faster rate during the early evolution. The system expands predominantly following the underlying Bjorken flow, which is reflected in the power-law type cooling observed in these results, except for a subtle decrease in the cooling rate around $\tau/\tau_R \sim 1$. This deviation occurs when collisional effects become comparable to the expansion rate, marking the characteristic thermalization timescale $\tau_R$. At this transition point, the collision term in the Boltzmann equation begins to compete effectively with the expansion term, resulting in a modification of the cooling behavior. One can also observe that the cooling rate is sensitive to the underlying particle statistics, reflecting the statistics-dependent coupling between the temperature 
$T$ and $\alpha$ through the Landau matching conditions. 


Next, we examine the time evolution of the normalized trace of the energy-momentum tensor $\Theta^{\mu}_{\;\mu}/T^4=\frac{m^2}{T^4}\rho_{-1,0}$. This is one of the primary observables of this study. Since we consider a constant temperature-independent mass, the normalized trace depends solely on the non-equilibrium distribution function through the moment $\rho_{-1,0}$ and its temporal evolution.

\begin{figure}[h]
    \centering
    \hspace*{-5.5mm}
    \includegraphics[width=1.07\linewidth]{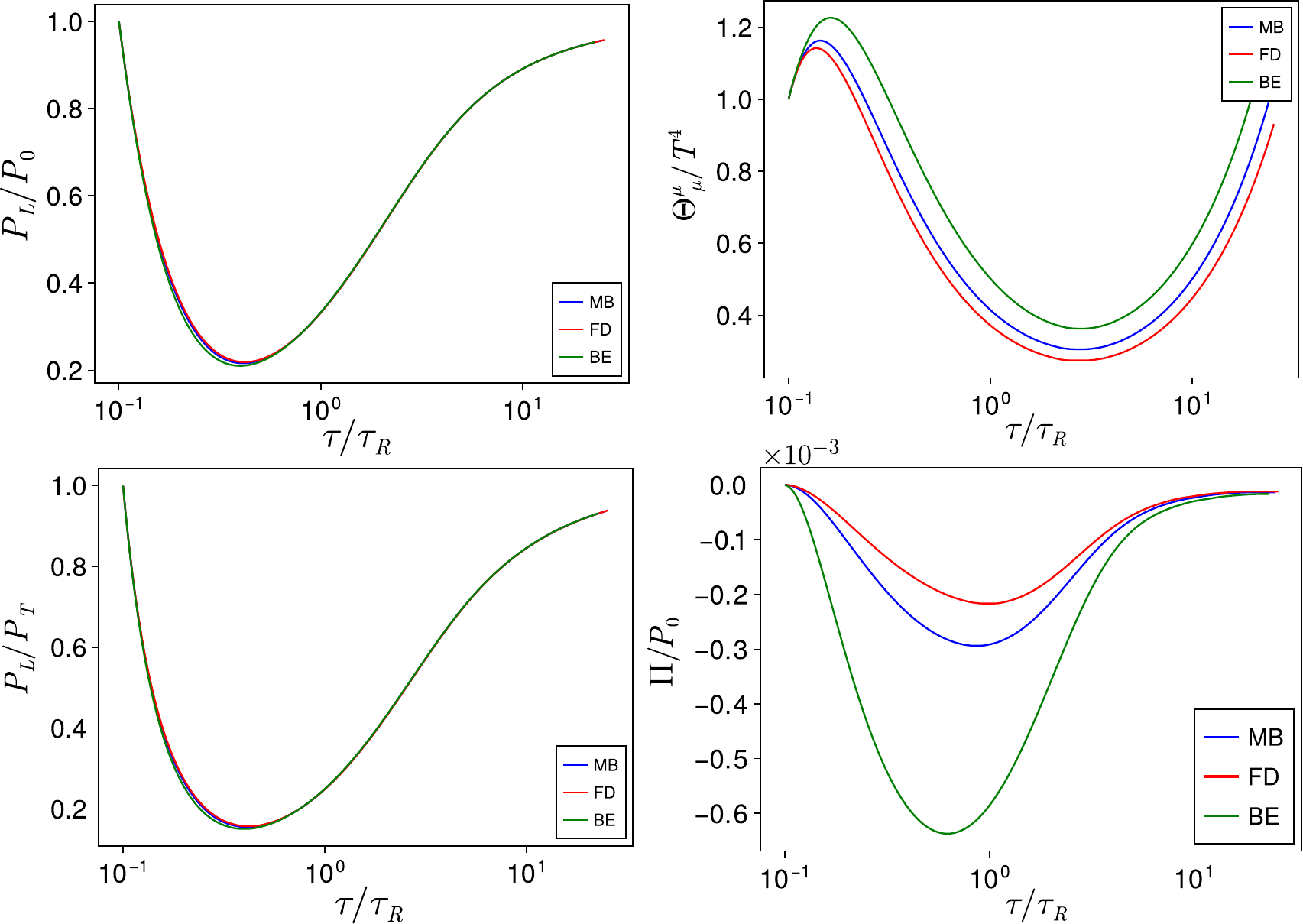}
    \caption{Time evolution of the $\Theta^{\mu}_{\; \mu}/T^{4}$ scaled to its initial value, the scaled bulk pressure, the scaled longitudinal pressure, and the ratio $P_{L}/P_{T}$ for Maxwell-Boltzmann, Fermi-Dirac, and Bose-Einstein statistics, starting from an equilibrium initial condition ($\rho_{n,l}=\rho^{eq}_{n,l}$) with $T_{0}=1~\mathrm{GeV}$ and $\alpha_{0}=0.0$.}
    \label{fig:NormTraceSing}
\end{figure}
In Fig.~\ref{fig:NormTraceSing}, we show $P_{L}/P_{0}$, $\Theta^{\mu}_{\;\mu}/T^4$, $P_{L}/P_{T}$, and $\Pi/P_{0}$ as functions of scaled time for different distributions, all starting from the initial values $\alpha_0 = 0.0$ and $T_0 = 1~\mathrm{GeV}$. Since we start from an equilibrium distribution (i.e., $\rho_{n,l} = \rho^{eq}_{n,l}$), the bulk pressure ($\Pi$) vanishes at initialisation. At early times, rapid longitudinal expansion drives the system out of equilibrium, causing $\Pi/P_{0}$ to develop non-zero values. These dissipative pressures reach their maximum magnitudes around $\tau \sim \tau_R$, the characteristic time when expansion and collision rates become comparable. At late times $\tau \gg \tau_R$, collisions dominate and drive the system back toward equilibrium, causing both $\Pi$  relax back to zero, as shown in Fig.~2.
Based on the different statistics, we do not observe any significant differences in $P_{L}/P_{T}$ and $P_{L}/P_{0}$ during the evolution; however, we observe a noticeable difference in the scaled bulk pressure, $\Pi/P_{0}$, for different statistics. 
As shown in Fig.~2 we observe that the ordering of scaled bulk pressure is 
$|\Pi_{BE}/P_{0}| > |\Pi_{MB}/P_{0}| > |\Pi_{FD}/P_{0}|$. 

Since we are starting from fixed initial values of $T$ and $\alpha$, this leads to different starting magnitudes of $\Theta^{\mu}{}_{\mu}/T^{4}$ for the three particle statistics. To enable direct comparison of the subsequent dynamical evolution among the MB, FD, and BE cases, we scale $\Theta^{\mu}{}_{\mu}/T^{4}$ by its value at $\tau=\tau_{0}$. For all three distributions, one can observe it exhibits non-monotonic behaviour: it shows a local maximum at very early times, followed by a dip around $\tau \sim \tau_R$, and then increases monotonically afterward within the range $\tau/\tau_R$ considered here. The early-time peak arises from the interplay between the decreasing temperature and the evolving moment $\rho_{-1,0}$. The subsequent dip around $\tau \sim \tau_{R}$ reflects the transition regime where collisional effects begin to dominate over expansion, temporarily reducing the trace before it rises again at later times. This qualitative behaviour is similar to that reported in \cite{PhysRevC.105.024911}.
\begin{figure}[h]
    \centering
    \hspace*{-2.5mm}
    \includegraphics[width=1.05\linewidth]{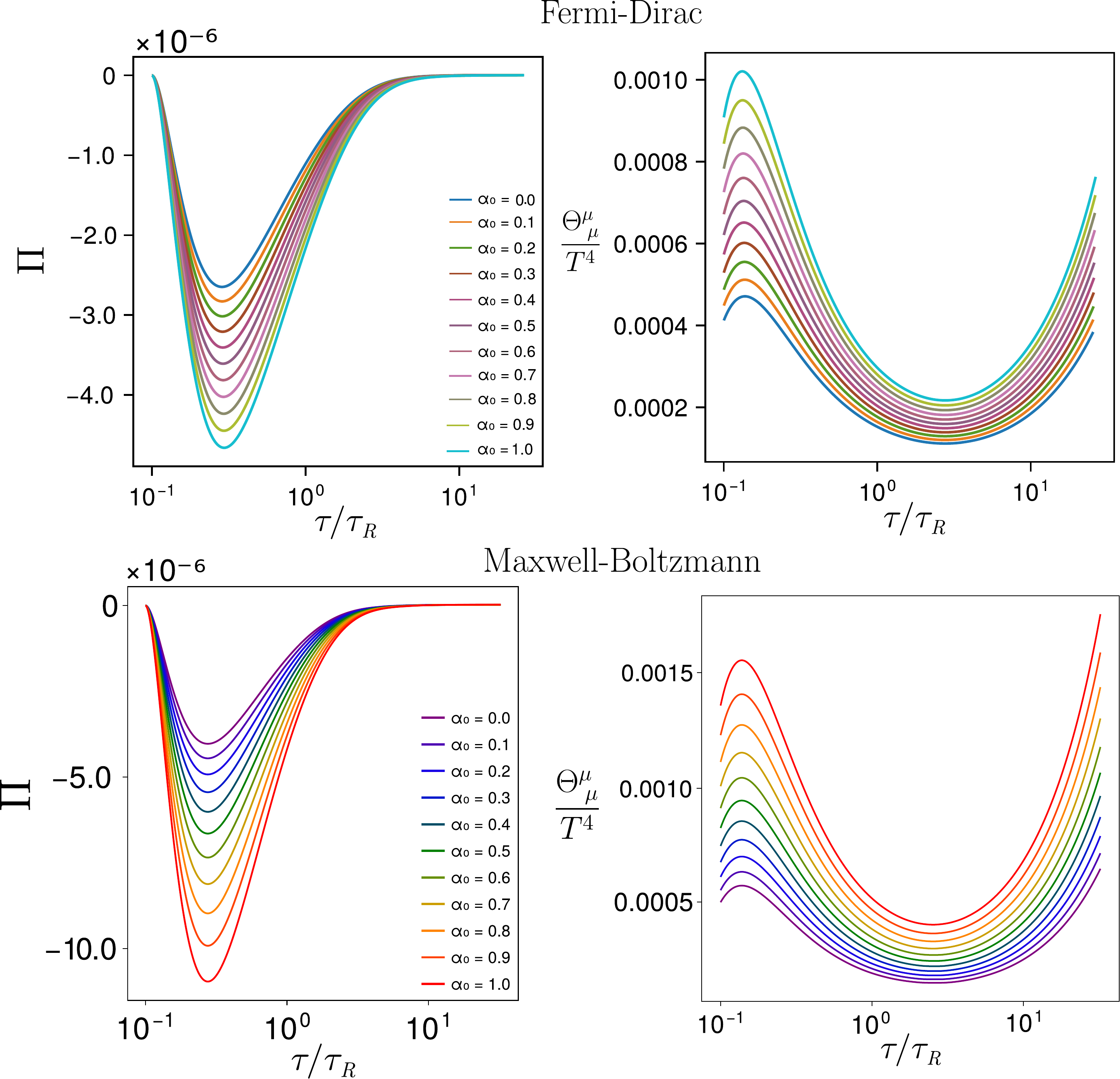}
    \caption{Evolution of bulk pressure ($\Pi$) and  normalized trace of the energy-momentum tensor  ($\Theta^{\mu}_{\;\mu}/T^{4}$) starting from  different initial chemical potentials.}
    \label{fig:placeholder_3}
\end{figure}

However, an important caveat must be noted. During the time evolution, both the temperature and chemical potential vary as shown in Fig.1. Therefore, our results cannot be directly compared with equilibrium lattice QCD calculations, which are typically reported for a constant chemical potential. In equilibrium lattice QCD studies, $\Theta^{\mu}_{\;\mu}/T^4$ exhibits a peak near the crossover temperature $T_c$ and decreases monotonically for both higher and lower temperatures \cite{BAZAVOV2014867}. This temperature dependence is intimately connected to the behavior of bulk viscosity, which also shows a peak near $T_c$ due to the rapid change in the equation of state. In contrast, our non-equilibrium calculation captures the time-dependent evolution of the trace, where both $T$ and $\mu$ are dynamically evolving.

Next, we examine the time evolution of the bulk pressure ($\Pi$) and normalized trace of the energy momentum tensor  ($\Theta^{\mu}_{\; \mu}/T^{4}$) by evolving the system from a range of different initial $\alpha$ values. Here also, we must keep in mind that all results shown above are for both $T$ and $\mu$ are dynamically evolving. 
In Fig.~\ref{fig:placeholder_3} we observe that the magnitude of the bulk and normalised trace of the energy-momentum tensor increases with increasing initial chemical potential $\mu_{0}$ (or equivalently $\alpha_0 = \mu_0/T_0$).
This trend can be understood physically by noticing that a larger chemical potential corresponds to a higher particle density, which enhances the deviation from equilibrium during the expansion-dominated phase. 
Furthermore, systems with higher initial chemical potential take longer to thermalize, as evidenced by the broader peaks $\Pi$ and delayed return to equilibrium at late times.

\begin{figure}[h]
    \centering
    \includegraphics[width=0.85\linewidth]{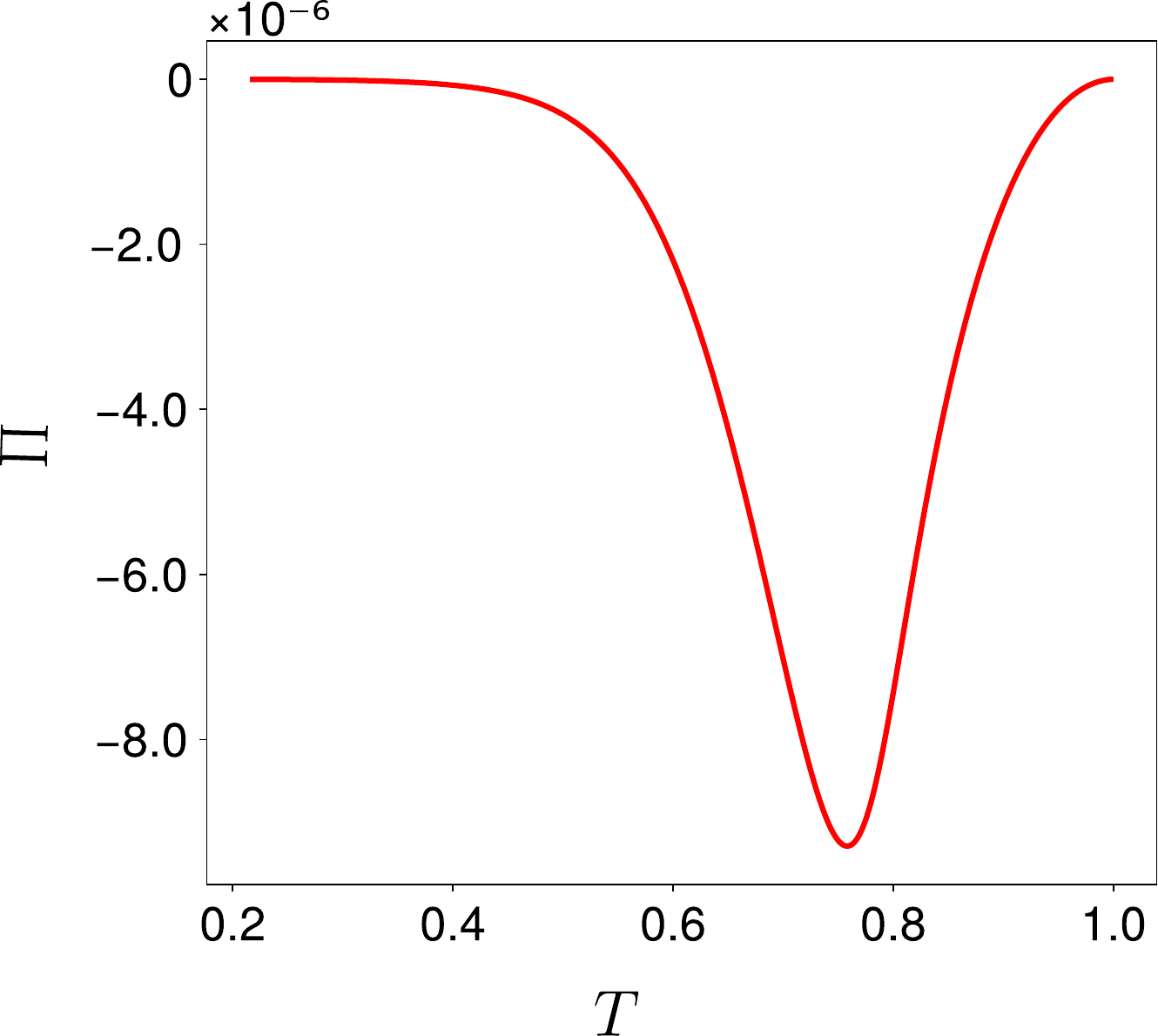}
    \caption{Bulk pressure $\Pi$ (for MB) as a function of temperature $T$, obtained by slicing the three-dimensional $(\alpha,T,\Pi)$ surface along $\alpha = 0$.}
    \label{fig:BulkStress}
\end{figure}
Bulk viscosity ($\zeta$) is an important input in the study of heavy-ion collisions, particularly in hydrodynamic models, since it influences the expansion rate of the fluid produced in these collisions. In hydrodynamic model studies, the bulk viscosity to entropy density ratio ($\zeta/s$) is typically taken either as a constant or, more realistically, as a temperature-dependent quantity. The bulk viscous coefficient governs the evolution of the bulk stress through a relaxation-type equation. Since we can directly compute the bulk stress within our formalism, it is instructive to study its temperature dependence at constant $\alpha$.

To characterize the bulk stress at zero chemical potential, we first compute the three-dimensional surface $\Pi(T,\alpha)$ by scanning the $T$-$\alpha$ plane for a range of initial conditions with $-1.0 < \mu_0 < 4.0$ GeV. We then extract the $\alpha = 0$ (i.e., $\mu = 0$) slice using the spline interpolation method to obtain the desired temperature dependence
\begin{equation}
\Pi(T) \equiv \Pi(T,\alpha)\big|_{\alpha=0}.
\end{equation}
The corresponding result is shown in Fig.~\ref{fig:BulkStress}.

We observe that, starting from a zero value (corresponding to the equilibrium initial condition), the bulk pressure rapidly develops a negative minimum at early times (high temperatures) and then asymptotically approaches the equilibrium value of zero from below in the late-time or low-temperature regime. This behavior is consistent with the relaxation-type evolution equation for bulk stress, where the relaxation time scale is controlled by the underlying microscopic collision time $\tau_R$. 

The negative bulk pressure at early times reflects the fact that the longitudinal expansion leads to non-equilibrium trace of the energy-momentum tensor which is less than its equilibrium value, resulting into $\Pi < 0$. As the system evolves and collisions restore isotropy, the bulk stress relaxes back toward zero. The early-time values of bulk stress are particularly important because the momentum-space anisotropy that gives rise to final-state flow harmonics is most strongly influenced by these non-zero dissipative effects during the initial stages of evolution. The magnitude and duration of the bulk stress directly affect the build-up of collective flow, making accurate determination of $\zeta(T)$ crucial for quantitative hydrodynamic modeling of heavy-ion collisions.
\subsubsection*{Non-equilibrium initial conditions}
In all the previous cases, we started from equilibrium initial conditions ($\rho_{n,l}=\rho^{eq}_{n,l}$). However, it is expected that the system immediately after the collision is far from equilibrium. Therefore, a more realistic choice for initial conditions would be non-equilibrium configurations. We now examine non-equilibrium initial conditions characterized by $\rho_{n,l}\neq \rho^{eq}_{n,l}$, except for $\rho_{0,0}$ and $\rho_{1,0}$, which must remain fixed to the equilibrium value to satisfy the Landau matching conditions. 

To generate non-equilibrium initial conditions, we perturb the moment ratios according to 
\begin{equation}
\rho_{n,l} = \rho^{eq}_{n,l}(1.0 + \delta\rho_{n,l}),
\end{equation}
where $\delta\rho_{n,l}$ is drawn from a Gaussian distribution $\mathcal{N}(0,\sigma^2)$ with zero mean and standard deviation $\sigma = 0.3$. This random perturbation scheme allows us to explore a representative ensemble of initial momentum distributions that deviate from equilibrium while still satisfying the energy and particle number constraints and other physical constraints given in \eqref{eq:Constraint_1}, \eqref{eq:Constraint_2}. We perform the simulation for thirty randomly selected initial configurations to assess the typical behavior and spread of possible evolution trajectories.

For all simulations in this subsection, we use an initial temperature of $T_{0} = 1$ GeV, mass $m = 0.1$ GeV, and scaled chemical potential $\alpha_0 = 0.0$.

\begin{figure*}[t!]
    \centering
    \includegraphics[width=0.9\linewidth]{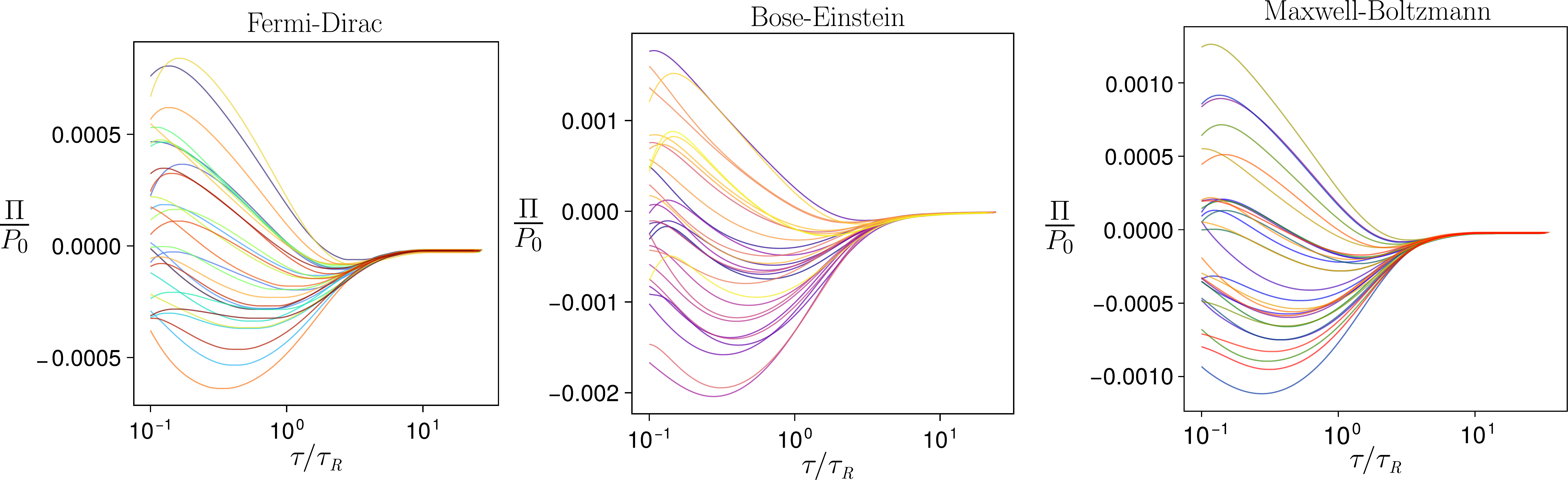}
    \caption{Time evolution of the scaled bulk  pressure $\Pi/P_{0}$ for three different distribution starting from non-equilibrium initial conditions ($\rho_{n,l} \neq \rho^{eq}_{n,l}$) for thirty randomly selected initial configurations. The spread of trajectories reflects the variation in initial momentum-space anisotropies.}
    \label{fig:ScaledBulkNonEq}
\end{figure*}
\begin{figure*}[t!]
    \centering   \includegraphics[width=0.9\textwidth]{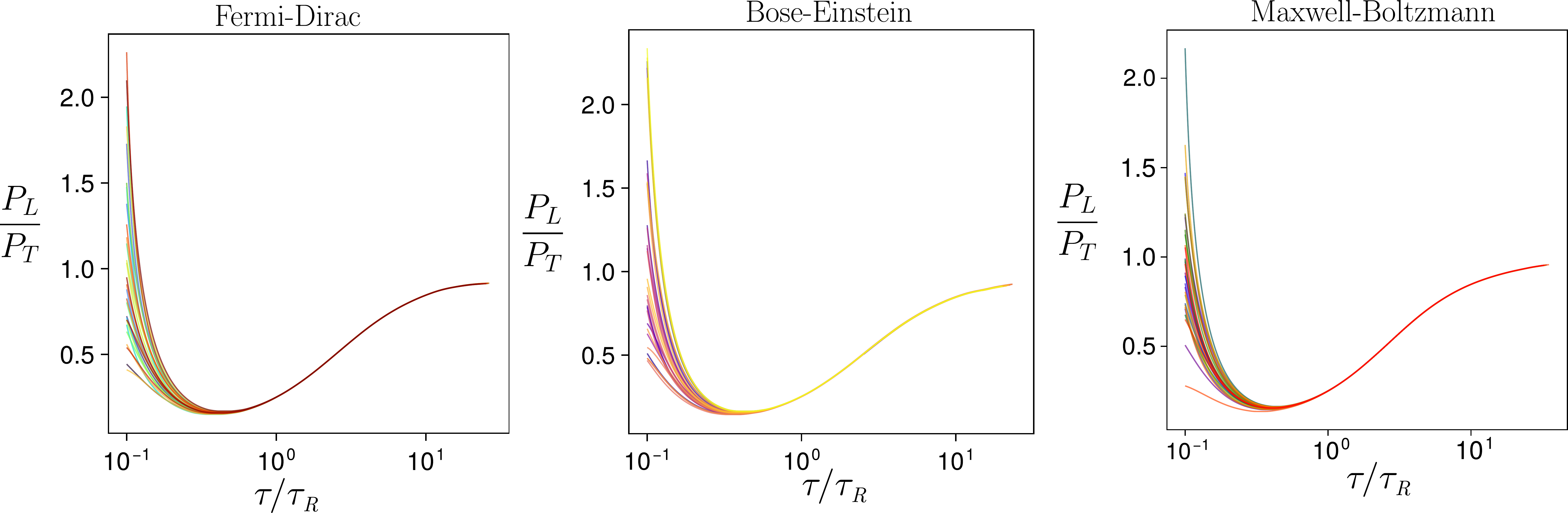}
    \caption{Time evolution of the pressure anisotropy ratio $P_{L}/P_{T}$ starting from non-equilibrium initial conditions ($\rho_{n,l} \neq \rho^{eq}_{n,l}$) for thirty randomly selected configurations for three different distributions. The approach to unity at late times indicates the restoration of pressure isotropy.}
    \label{fig:PLPTNonEq}
\end{figure*}

\begin{figure*}[t!]
    \centering
    \includegraphics[width=0.9\linewidth]{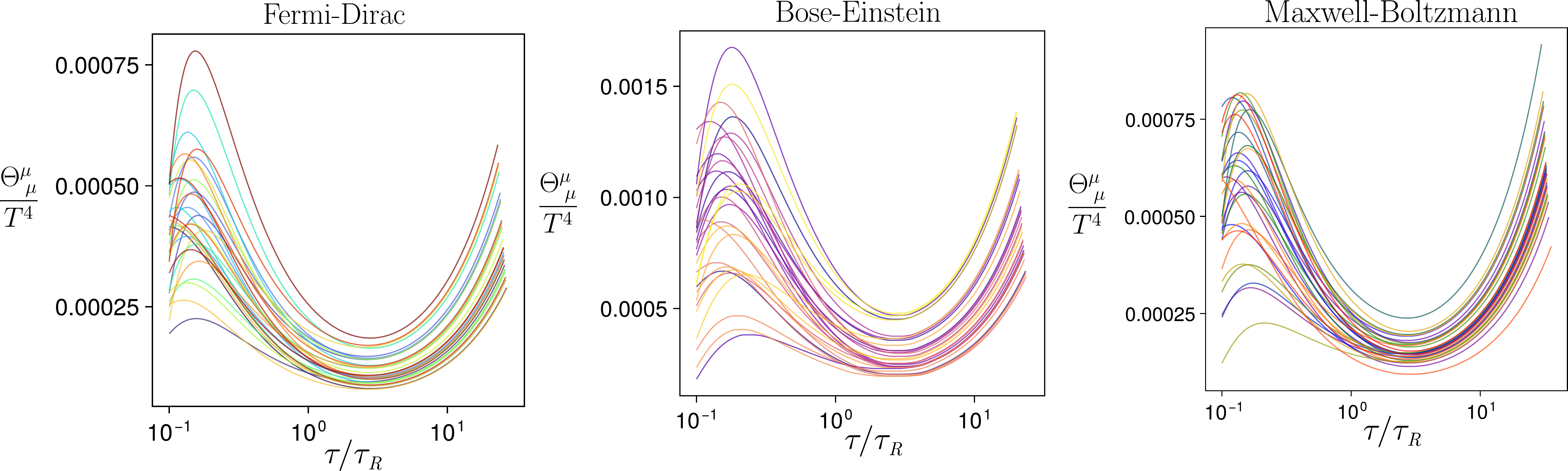}
    \caption{Time evolution of the normalised trace $\Theta^{\mu}_{\; \mu}/{T^4}$ starting from non-equilibrium initial conditions ($\rho_{n,l} \neq \rho^{eq}_{n,l}$) for randomly selected initial configurations.}
    \label{fig:NormTraceMult}
\end{figure*}

The behavior of the scaled bulk pressure $\Pi/P_{0}$ is shown in Fig.~\ref{fig:ScaledBulkNonEq}. Starting from non-zero initial values that depend on the specific realization of the perturbations, all trajectories evolve toward equilibrium as $\tau/\tau_{R}$ increases, with the scaled bulk pressure asymptotically approaching zero at late times. The spread among different trajectories at early times reflects the variation in initial momentum-space configurations, while the convergence at late times demonstrates the robustness of the thermalization/isotropisation process governed by the collision term. The fact that all trajectories, regardless of their initial deviations from equilibrium, eventually relax toward zero bulk pressure confirms the stability of the equilibrium state and the efficacy of the relaxation time approximation in driving the system toward local thermalization. 

A similar trend is observed in the pressure anisotropy ratio $P_{L}/P_{T}$ shown in Fig.~\ref{fig:PLPTNonEq}. In equilibrium, the pressure is isotropic and $P_{L}/P_{T} = 1$. Starting from non-equilibrium initial conditions where the longitudinal and transverse pressures can differ significantly (reflecting momentum-space anisotropies), the system gradually evolves toward isotropy. All trajectories converge to $P_{L}/P_{T} \to 1$ at late times, indicating that the system reaches pressure isotropy as it approaches local equilibrium, irrespective of the particle statistics. The rate of convergence is controlled by the relaxation time $\tau_R$, with most trajectories showing significant progress toward isotropy by $\tau/\tau_R \sim 0.5-1$, a much faster rate than the bulk-pressure relaxation time.

Finally, we examine the non-equilibrium trace $\Theta^{\mu}_{\;\mu}/T^4$ as a function of $\tau/\tau_{R}$ in Fig.~\ref{fig:NormTraceMult}. The normalized trace exhibits more complex behavior compared to the bulk pressure and pressure anisotropy, showing non-monotonic evolution with structures that depend on the initial conditions. Although the early-time evolution of $\Theta^{\mu}{}{\mu}/T^{4}$ exhibits no systematic ordering across different initial conditions, the trajectories converge to a well-defined ordering for $\tau/\tau{R} > 1$.  Despite the variation in trajectories at early and intermediate times, all configurations eventually evolve toward a common late-time behaviour. However, the persistent differences among trajectories even at relatively late times (compared to $\Pi/P_{0}$ and $P_L/P_T$) suggest that the trace is more sensitive to the detailed structure of the momentum distribution.

\section{Conclusion}
In this work, we investigated the far-from-equilibrium dynamics of a massive relativistic Boltzmann gas undergoing boost-invariant Bjorken expansion by solving the RTA moment equation with the method of moments. Introducing an infinite hierarchy of moments $\rho_{n,l}$ along with the Landau matching condition for energy and particle density. From matching conditions we get evolution for $T$ and $\alpha$ and all other non-equilibrium moments by solving the coupled differential equations. This further allow us to compute the trace anomaly $\Theta^{\mu}_{\; \mu} = \varepsilon - 3P$, longitudinal, transverse pressure, and bulk viscous pressure for a non-conformal system with a finite mass.

We showed that when we start from an equilibrium initial condition, both $T$, $\alpha$ decrease with proper time with a characteristic change around $\tau \sim \tau_{R}$, where $\tau_{R}$ is the relaxation time. It means that the characteristic change took place when the microscopic and macroscopic time scales became comparable. The time evolution of $T$ and $\alpha$ depends on the particle statistics. The normalized trace anomaly $\Theta^{\mu}_{\; \mu}/T^{4}$ exhibits a non-monotonic time dependence, characterized by an early maximum, a dip near $\tau \sim \tau_{R}$, and a subsequent late-time increase, with an overall magnitude that grows with the initial chemical potential. The bulk pressure $\Pi$ develops dynamically due to rapid longitudinal expansion and hits the extrema around $\tau \sim \tau_{R}$, and relax back to zero. The value of $\Pi/P_{0}$ exhibits a clear dependence on the underlying particle statistics.  We found that the increasing initial chemical potential amplifies the extrema of $\Pi$ and delays the relaxation which indicates that the system with higher chemical potential remains out of equilibrium for a longer amount of time. To characterise the behaviour of bulk pressure as a function of temperature at vanishing chemical potential($\mu = 0$), we extracted the $\Pi(T)$ at $\alpha = 0$, which shows that the bulk pressure attains an extrema before it slowly approaches zero as the system cools. Similarly, $\Theta^{\mu}_{\; \mu}/T^{4}$ increases with increasing chemical potential. This behaviour is consistent with the relaxation type description of the bulk viscous pressure and encodes information about the effective bulk viscosity in a dynamical setting rather than assuming its equilibrium value.

We then considered a more realistic class of initial conditions where the system starts far from equilibrium, with all moments randomly perturbed around $\rho_{n,l}=\rho^{eq}_{n,l}$ while keeping $\rho_{0,0}= \rho^{eq}_{0,0}$ and $\rho_{1,0}=\rho^{eq}_{1,0}$ to satisfy Landau matching. 
For an ensemble of such non-equilibrium initial states, we have shown that the scaled bulk pressure $\Pi/P_{0}$ and the ratio of $P_{L}/P_{T}$ converge to a common late-time trajectory, exhibiting attractor-like behaviour. This attractor-like behaviour shows the equilibration in RTA kinetic theory and provides a microscopic realisation of bulk and shear hydrodynamic attractors in a non-conformal system. On the other hand, $\Theta^{\mu}_{\; \mu}/T^4$
 exhibits a noticeable spread among different trajectories even at late times within the time range considered here.  This indicates that the trace anomaly remains sensitive to the evolution history and retains information about the early-time dynamics over a longer period. Consequently, non-equilibrium corrections to the equation of state require particular care in their treatment.


\begin{acknowledgments}
V.R. acknowledges the financial support from Anusandhan National Research Foundation (ANRF), India through Core Research Grant , CRG/2023/001309. K.S, R.G., and V.R. also acknowledges Department of Atomic Energy (DAE), India for financial support.
\end{acknowledgments}



\newpage
\twocolumngrid     
\bibliography{References}

\end{document}